\documentclass[aps,final,notitlepage,oneside,twocolumn,nobibnotes,nofootinbib,
superscriptaddress,noshowpacs,centertags]{revtex4-1}

\usepackage[utf8]{inputenc}
\usepackage[english]{babel}
\usepackage{graphicx}
\usepackage{latexsym}
\usepackage{amssymb}
\usepackage{amsmath}
\usepackage{float}
\usepackage{color}

\begin{document}

\title{
Origin of monolithic high-$z$ galaxies and UV luminosity of mergering high-$z$ galaxies in the cosmological model with non-standard spectrum of density perturbations
}
	\author{Yu.N. Eroshenko}\thanks{e-mail: eroshenko@inr.ac.ru}
	\affiliation{Institute for Nuclear Research of the Russian Academy of Sciences,
	60th Anniversary of October Prospect 7a, 117312 Moscow, Russia}
	\author{V.N. Lukash}\thanks{e-mail: lukash@asc.rssi.ru}
	\affiliation{Astro Space Center, Lebedev Physical Institute of the Russian Academy of Sciences,
	Profsoyuznaya str.\ 84/32, 117997 Moscow, Russia}
	\author{E.V. Mikheeva}\thanks{e-mail: helen@asc.rssi.ru}
	\affiliation{Astro Space Center, Lebedev Physical Institute of the Russian Academy of Sciences,
	Profsoyuznaya str.\ 84/32, 117997 Moscow, Russia}
	\author{S.V. Pilipenko}\thanks{e-mail: spilipenko@asc.rssi.ru}
	\affiliation{Astro Space Center, Lebedev Physical Institute of the Russian Academy of Sciences,
	Profsoyuznaya str.\ 84/32, 117997 Moscow, Russia}
	\author{M.V. Tkachev}\thanks{e-mail: mtkachev@asc.rssi.ru}
	\affiliation{Astro Space Center, Lebedev Physical Institute of the Russian Academy of Sciences,
	Profsoyuznaya str.\ 84/32, 117997 Moscow, Russia}

	\date{\today}

\begin{abstract}
The James Webb Space Telescope (JWST) has detected an unexpectedly large number of galaxies at redshifts $z\geq 10$ compared to the predictions of the standard $\Lambda$CDM model. One of possible explanations is the presence of an excess (bump) in the power spectrum of perturbations at the mass scale of these galaxies, which is about $10^{10}M_\odot$. This excess simultaneously shifts the epoch of cosmic reionization to significantly earlier times, in contradiction with observations. Here we show that this defect of the bump model can be avoided if the perturbation spectrum has a cutoff (suppression) at smaller scales, which can be realized by lowering the amplitude of the primordial perturbation spectrum or by considering warm dark matter. With a cutoff present, fewer low-mass halos form and, consequently, fewer stars producing ionizing UV radiation. We also derive the halo merger-rate distribution in the presence of a bump using the extended Press--Schechter (EPS) theory, and verify this distribution against direct $N$-body simulations. Based on this distribution, and under the assumption that mergers trigger starbursts, we compute the UV luminosity function of early galaxies and demonstrate its agreement with available observational data. In the model under consideration, the fraction of early galaxies forming via the monolithic mechanism (through a single large-scale collapse) is significantly increased in comparison with the standard $\Lambda$CDM model, and the first stars appear directly in halos with masses $\geq10^8M_\odot$. We refer to such stars with primordial chemical composition as ``Population~IV stars,'' to distinguish them from the evolutionary different stellar populations.
\end{abstract}

\maketitle

%%%%%%%%%%%%%%%%%%
\section{Introduction}
%%%%%%%%%%%%%%%%%%

The development of observational cosmology over the past two decades has significantly improved the precision of cosmological parameters determination. A joint analysis of data from Planck and BICEP/Keck \cite{BICEPlanck} substantially weakened the viability of many inflationary models, since the measured local tilt of the density perturbation spectrum $n_{0.05}=0.965\pm 0.006$ and the relative amplitude of cosmological gravitational waves $r_{0.05}<0.035\,(2\sigma)$ (both evaluated at the pivot scale $k=0.05$~Mpc$^{-1}$) are compatible with only a few models (see the discussion in \cite{linde2025, cascade}) and/or require abandoning General Relativity in favor of, e.g., scalar-tensor theories of gravity \cite{teleparalel}. The measured value, $n_S=0.9671 \pm 0.0058$, is supported by contemporary CMB experiments, such as ACT (Atacama Cosmology Telescope) and SPT (South Pole Telescope) \cite{SPT}. At the same time, when baryon acoustic oscillation data from DESI (Dark Energy Spectroscopic Instrument) are included in the joint analysis, the spectrum turns out to be significantly closer to scale-invariant, $n=0.974\pm 0.003$ \cite{ACT}. These measurements concern the perturbation spectrum at sufficiently large scales. However, on galactic and sub-galactic scales the power spectrum is not yet precisely known, and various shapes are admissible, including a non-trivial spectrum with some features such as breaks and additional enhancements or suppressions. 

The deviation of the spectral tilt from unity and the negligibly small ratio of the gravitational-wave perturbation amplitude to the scalar one are currently firmly-established facts, which are sufficient to reconstruct the inflaton potential directly from large-scale observations \cite{cascade}. The small-scale regime ($<1$~Mpc) contains information about the finer details of the potential shape, namely, its additional parameters, which leave no observable imprint at large scales.

Since 2023, results of JWST high-redshift ($z>5$) galaxy samples have been actively published \cite{Naidu2022b, Castellano22, Finkelstein22, Donnan23, Labbe23}. Recent observations show that supermassive black holes in centers of early galaxies are considerably more massive than predicted by the standard $\Lambda$CDM model, and this excess is not due to selection effects (see \cite{Li_2025_ApJ_981} and references therein). Theoretical models of the early galaxies and early black hole formation thus face the challenge of explaining these observations.

In this work we consider cosmological models characterized by both an excess of power in the primordial power spectrum at sub-Mpc scales, which is favorable from the JWST perspective for the formation of halos with masses $\sim10^{10} M_\odot$, and a suppression (cutoff) at smaller scales ($\le 10^8 M_\odot$ in terms of halo masses), which prevents early star formation and early reionization. The point is that an excess of power, whether due to a bump or an enhanced spectrum tilt, inevitably leads to the early production of not only dark matter (DM) halos but also stars that efficiently ionize the interstellar and intergalactic medium. As a result, cosmological models with a power excess caused by a steeper small-scale spectrum contradict the Planck results on reionization \cite{zref}. A small-scale cutoff can not only resolve this problem but also reduce the number of low-mass satellite halos around massive galaxies.

A suppression of the power spectrum on small scales may arise not only from the shape of the primordial spectrum, but also from the presence of warm dark matter (WDM), which has been studied in various contexts for more than 20 years. In that case, small-scale perturbations are suppressed by residual thermal velocities of the DM particles. We thus build a model in which the first stars form not in low-mass but in massive halos, so that the hydrogen cooling mechanism is bremsstrahlung (free-free emission). 
We call the resulting stars as Population~IV stars to distinguish them from Population~III stars, which form in low-mass halos. These stars, rather than Population~III stars, may have been responsible for the initial metal enrichment of the Universe.

Galaxy formation models have to satisfy the full set of available observational data and constraints. We have already mentioned the requirement to reproduce the observed excess of galaxies at high $z$. In our model this effect is reproduced via a bump at some mass scale, as was shown in earlier works \cite{Tkaetal23, Eroshenkoetal2024, TkachevetalPRD2024, TkachevetalPRD2025, Eroshenko2025}. As those investigations showed \cite{Eroshenko2025}, bump spectra have a radical effect on the sky-averaged 21~cm neutral-hydrogen line profile. A bump on the scale of dwarf galaxies was also considered in \cite{Qinetal25}, although that work focused primarily on the formation of supermassive black holes.

A non-standard spectrum supplemented by a bump and a cutoff not only reduces ionization at high redshifts but also supports the so-called monolithic mechanism of galaxy formation, what means the formation of massive halos without hierarchical clustering of DM \cite{coldcollapse1999} and supported by recent observations \cite{McGetal24, Sil25}.

Another set of observational data that we reproduce in our model is the UV luminosity function of galaxies, which in recent years has become a useful tool for probing the early Universe \cite{FerPalDay23, Blaetal25, Peretal25, Aaretal25}. The Hubble Space Telescope (HST) has mapped the UV luminosity function out to $z\simeq10$, and the JWST has extended the explored range to $z\simeq17$.

In this paper we develop, within the EPS formalism \cite{Bonetal91, LacCol93}, a method for computing the distribution of DM halos by mass accretion rate associated with merging smaller halos by larger ones, which allows us to compute the UV luminosity function. This method differs from previously proposed approaches to halo mass growth, where the growth of a halo with a fixed initial mass was tracked. In our approach, mergers of halos with different masses are accounted for under the condition that the mass accretion rate on the larger halo is specified, regardless of its initial mass. Comparisons of this method with existing simulation results for the standard  featureless spectrum and with new simulations, performed for a spectrum with a bump (but without a cutoff), show good quantitative agreement. Each merger between halos of different masses triggers a starburst and the UV flux, whose characteristics can be specified within semi-analytic approximations. As we will show, our derived accretion-rate distribution provides good agreement with the observed UV luminosity function.

Our goal is therefore to demonstrate that a model with a bump and cutoff in density perturbation spectrum (non-standard spectrum) satisfies the following observational criteria:
\begin{itemize}
    \item At redshift $z\sim 9$, massive galaxies ($\sim10^{10}M_\odot$) are observed to be roughly one order of magnitude more abundant than predicted by the standard $\Lambda$CDM model, and at $z\sim 12$ the discrepancy is approximately two orders of magnitude.
    \item The epoch of reionization is complete by $z\simeq 6$, and 50\% ionization occurred at $z\sim 8$.
    \item The model has to reproduce the UV luminosity function of early galaxies at least to $z\sim12$, where the observational data can be considered sufficiently reliable.
\end{itemize}

%%%%%%%%%%%%%%%%%%%%%%%%%%%%%%%%%%%%%%%%%%%%%%%%%%%%
\section{Power spectrum with a bump and a cutoff, and early formation of massive galaxy halos}
\label{spesec}
%%%%%%%%%%%%%%%%%%%%%%%%%%%%%%%%%%%%%%%%%%%%%%%%%%%%

The power spectrum with a bump is conveniently parameterized as
%(1)
	\begin{equation}
		\frac{P_{\rm bump}(k)}{P_{\Lambda CDM}(k)}=1+A\cdot\exp\left(-\frac{(\log(k)-\log(k_0))^2}{\sigma_k^2}\right),
		\label{bump}
	\end{equation}
where $P_{\Lambda CDM}(k)$ is the density perturbation power spectrum of the standard $\Lambda$CDM model, $k$ is the wavenumber, and $A$, $k_0$, and $\sigma_k$ are constants. We consider a cosmological model with $A=20$, $k_0=4.69$~Mpc$^{-1}$, and $\sigma_k=0.1$. Following the notation proposed in \cite{Tkaetal23}, we refer to this model as $gauss\_1$.

In contrast to the bump, whose origin can be naturally associated with processes occurring during an inflation, the cutoff in the power spectrum may arise either from a suppression in the primordial spectrum of perturbations or from the nature of DM.
In the latter case, the cutoff results from free-streaming of DM particles.
This process was considered, for example, in \cite{BerDokEro03}, where for the Fourier component of the particle number density one obtained
(2)
\begin{equation}
n_{k}(t) = n_{k}(t_d)
\exp\left(-\frac{1}{2}k^2g^2(t)\frac{T_d}{m_{\chi}}\right),
\label{nk}
\end{equation}
where $t_d$ and $T_d$ are the kinetic decoupling time of DM from radiation and the temperature at that moment, respectively, $m_{\chi}$ is the DM particle mass, and the function
(3)
\begin{equation}
g(t)=a(t_d)\int \limits_{t_d}^{t}\frac{dt'}{a^2(t')}
\label{defg}
\end{equation}
saturates near the matter-radiation equality epoch $t\sim t_{\rm eq}$, i.e., it approaches a constant for $t\gg t_{\rm eq}$.

In this work we do not specify a DM particle model and use only the general form of the free-streaming cutoff as an exponential, as in Eq.~(\ref{nk}). We characterize the cutoff position by the mass $M_{\rm fs}$ and the corresponding wavenumber $k_{\rm fs}$, so that this exponential takes the form $\exp(-k^2/k_{\rm fs}^2)$, equivalent to a factor $\exp(-M_{\rm fs}^{2/3}/M^{2/3})$ in the transfer function.

Thus, the full density perturbation power spectrum can be presented as
%(4)
\begin{equation}
P_{\rm fs}(k)=P_{\rm bump}(k)\exp(-2k^2/k_{\rm fs}^2),
\label{pfs}
\end{equation}
and the dispersion of the density perturbation is as follows:
%(5)
\begin{equation}
\sigma^2(M)=\frac{1}{2\pi^2}\int_0^\infty P(k)W^2(kR)\,k^2 dk,
\label{dispdef}
\end{equation}
where $P(k)$ denotes $P_{\Lambda CDM}$ for the standard $\Lambda$CDM model, $P_{\rm bump}(k)$ for the bump model, or $P_{\rm fs}(k)$ for the model with both a bump and a cutoff; $W(y)=3(\sin y -y\cos y)/y^3$ is the Fourier transform of the top-hat window function. The mass of a gravitationally bound object and the smoothing scale are related by $M=4\pi\bar \rho R^3/3$, where $\bar\rho$ is the mean matter density of the Universe.

\begin{figure}
\centering
\includegraphics[width=0.48\textwidth]{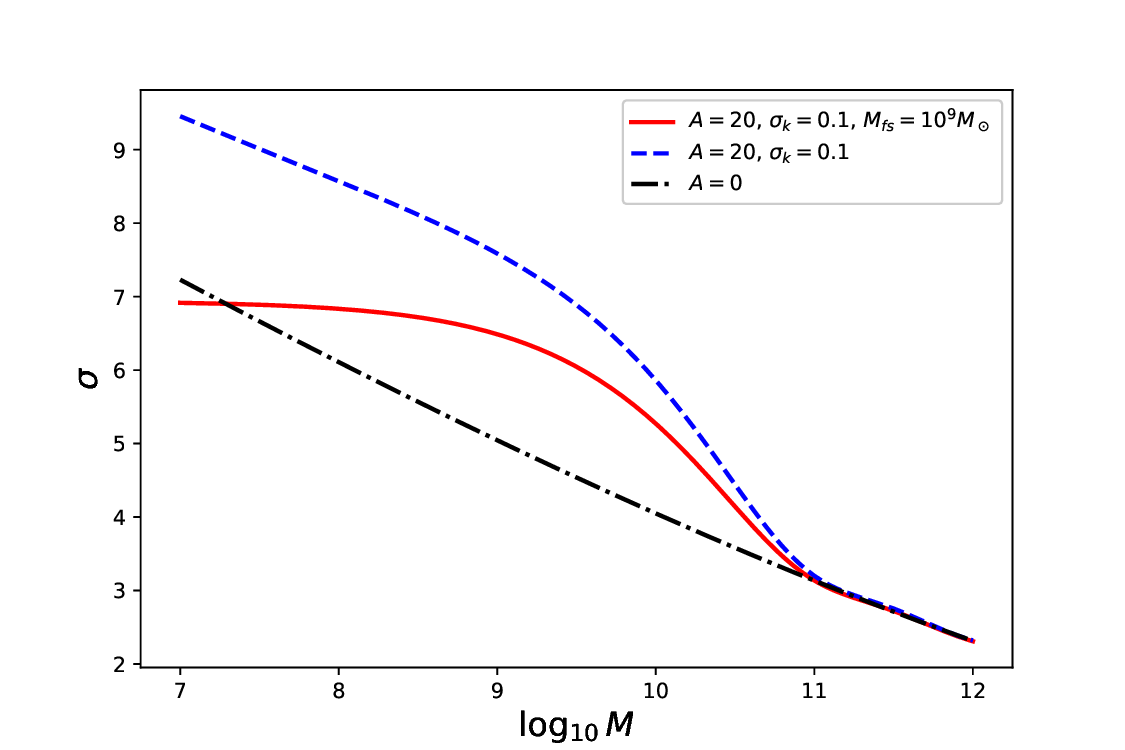}
\caption{Density perturbation dispersion in the models with a bump (blue dashed line), with a bump and a cutoff (solid red line), and in the standard $\Lambda$CDM model (black dot-dashed line). Parameter values are indicated in the legend.}
\label{gr1}
\end{figure}

Figure~\ref{gr1} illustrates the behavior of the dispersion in the models considered; the filtering is performed on the linear scale corresponding to $M_{\rm fs}=10^9M_\odot$. The black dot-dashed line corresponds to the standard $\Lambda$CDM model, the blue dashed line to the bump model, and the red solid line to the spectrum that contains both a bump and a small-scale cutoff.

The cutoff in the WDM power spectrum is often fitted by power-law functions \cite{BodOstTur01}, but the exact shape of the cutoff is not crucial for the purposes of this paper, since we also consider a variant with a cutoff in the primordial perturbation spectrum. We also note that \cite{BerDokEro03} considered massive ($\sim70$~GeV) particles (WIMPs) that underwent kinetic decoupling while already non-relativistic. In contrast, WDM models typically involve particles with masses of order keV that decoupled while still relativistic. For this case, \cite{BodOstTur01} found the relation between the characteristic scale and the particle mass: $k_{\rm fs}\sim 13(m_\chi/\mbox{keV})^{1.15}$~Mpc$^{-1}$.

%=================================Fig1
%================================================
%================================Fig2
\begin{figure}
\centering
\includegraphics[width=0.48\textwidth]{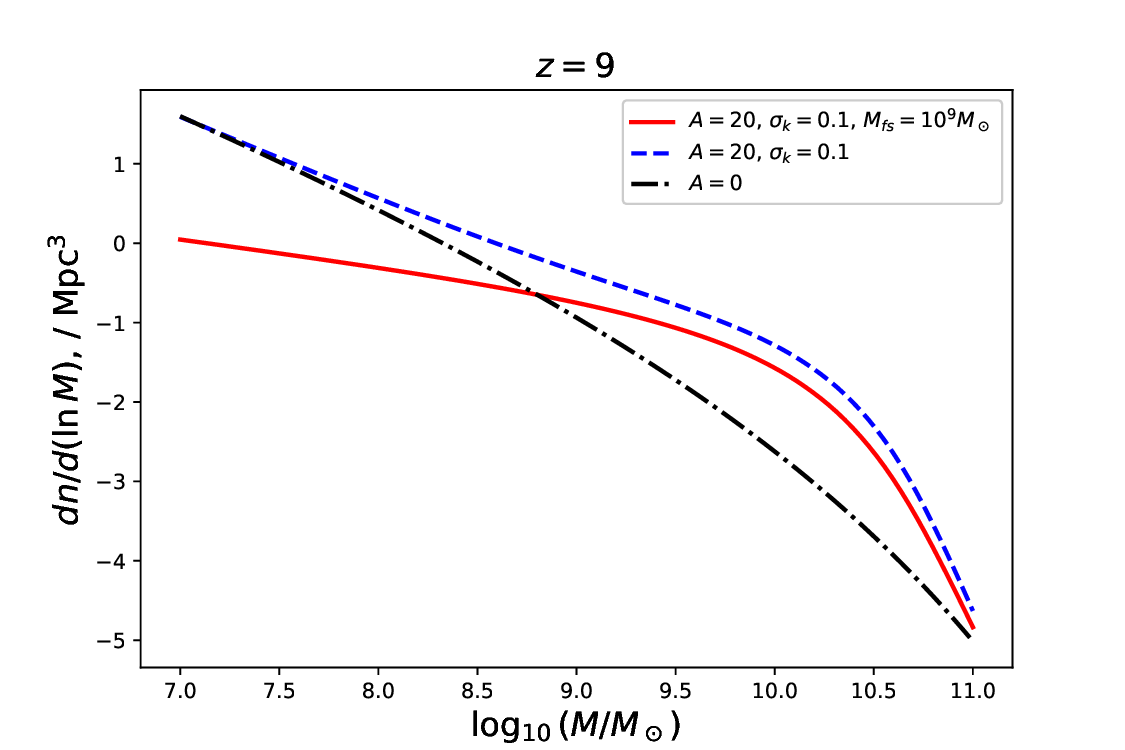}\\
\includegraphics[width=0.48\textwidth]{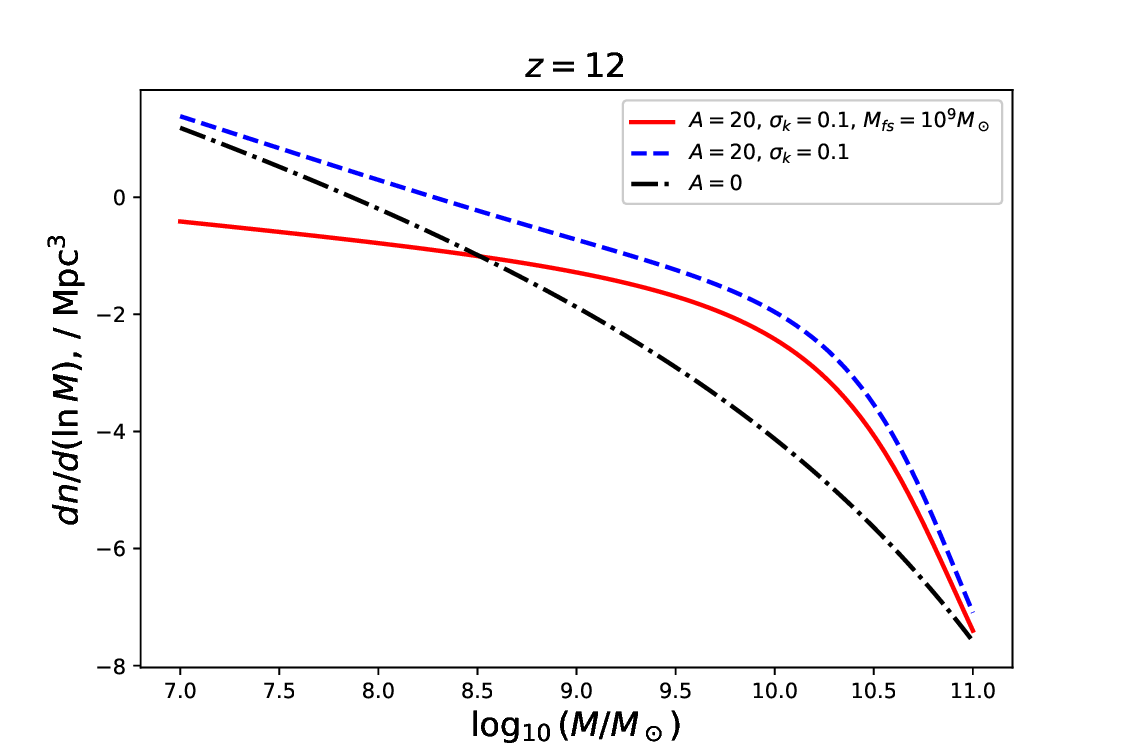}
\caption{Differential DM halo mass function computed from the EPS formalism (Sheth--Tormen formula \cite{ST}) at $z=9$ (top) and $z=12$ (bottom).}
\label{gr2}
\end{figure}
%====================================

Figure~\ref{gr2} illustrates the differential mass function of gravitationally bound halos calculated within the EPS formalism (Sheth--Tormen formula \cite{ST}) for the spectra considered in this work. As seen from the figure, the model with a bump and a cutoff satisfies the first observational criterion formulated in the Introduction. In addition, the suppression of the spectrum amplitude on small scales in the model with a cutoff is capable of resolving the satellite overabundance problem that exists in the standard $\Lambda$CDM model.

In this work we assume that the spectrum amplitude is suppressed at scales $M\leq10^9M_\odot$. This leads to significantly fewer low-mass galaxies forming compared to the standard model. At the same time, the presence of the bump causes earlier and more efficient formation of more massive galaxies. This means that a fraction of massive galaxies formed not hierarchically through mergers of smaller building blocks, but through a single large-scale collapse, which is known as monolithic mechanism. Observations do suggest that some fraction of galaxies may have formed via the monolithic mechanism.

It is worth noting that the monolithic collapse scenario historically predates the hierarchical model; only after the concept of cold DM became established did the picture of multiple successive mergers of low-mass protogalactic fragments emerge as the standard way of galaxy formation. The cooling of gas in a massive DM halo and the formation of a stellar galaxy was first studied in \cite{WhiRee78}. The condition for gas to collapse is the inequality $t_{\rm cool}<t_d$, where $t_{\rm cool}$ is the cooling time and $t_d$ is the dynamical time of the DM halo. The aforementioned condition determines the maximum mass of an object in which gas settles to the center and fragments into stars. If the condition is not satisfied, the gas remains distributed throughout the halo, with its temperature maintained at the halo virial temperature as in the case of galaxy clusters, which are filled throughout their volume with hot gas.

%%%%%%%%%%%%%%%%%%%%%%%%%%%%%%%%%%%%%%%%%%%%%%%%%%%%%%%%%%%%%%%%
\section{Evolution of the intergalactic gas ionization fraction}
%%%%%%%%%%%%%%%%%%%%%%%%%%%%%%%%%%%%%%%%%%%%%%%%%%%%%%%%%%%%%%%%

The evolution of the ionization fraction $x$ in the case of the standard density perturbation spectrum has been studied, for example, in \cite{Fur06, Furlanetto2006}. Analogous calculations can be performed for a model with a non-standard spectrum containing a bump and a cutoff. 

The ionization fraction $x$ evolves with time $t$ according to \cite{Fur06, Furlanetto2006}
%(6)
\begin{equation}
\frac{{\rm d}x}{{\rm d} t}=\zeta\frac{{\rm d} f_{\rm coll}[z,M_{\rm cooling}(z)]}{{\rm d} t}-\alpha_rx^2n_HC(z),
\label{ioneq}
\end{equation}
where $z$ is the redshift; $\zeta=A_{\rm He}f_*f_{\rm esc}N_{\rm ion}$ is the ionization efficiency parameter; 
$A_{\rm He}=1.22$ is the helium correction factor; 
$f_*$ is the fraction of baryons incorporated into stars (here $f_*=0.1$); 
$f_{\rm esc}$ is the fraction of ionizing photons that escape the galaxy into the intergalactic medium; 
$N_{\rm ion}$ is the number of ionizing photons per baryon ($N_{\rm ion}=0.45N_\alpha$ for Population~II stars, where $N_\alpha$ is the number of photons per baryon in the frequency range from the Ly$\alpha$ line to the Lyman limit); 
$f_{\rm coll}$ is the fraction of collapsed matter in units of the minimum halo mass $M_{\rm cooling}$ in which gas cooling via free-free transitions is possible; 
$n_H$ is the number density of hydrogen atoms;
$\alpha_r=4.2\times10^{-13}$~cm$^3$~s$^{-1}$ is the recombination coefficient; 
$C=\langle n_e^2\rangle/\langle n_e\rangle^2$ is the clumping factor; 
and $n_e$ is the electron density in the ionized medium.

Several remarks are in order regarding some of the quantities listed above.

Within EPS, the mass fraction of matter incorporated into virialized halos with masses $>M$ is
%(7)
\begin{equation}
f_{\rm coll}(z,M)={\rm erfc}\left\{\frac{\delta_c}{\sqrt{2}\sigma(M)D(z)}\right\},
\end{equation}
where $\delta_c$ is the critical overdensity for halo collapse ($\delta_c=1.69$ here), and $D(z)$ is the linear growth factor with $D(0)=1$.

The minimum DM halo mass, in which gas cooling via free-free transitions is possible, corresponds to a virial temperature of $\sim10^4$~K. This mass evolves with time and can be presented as a function of redshift:
%(8)
\begin{equation}
M_{\rm cooling}(z)=1.9\times10^7\left(\frac{1+z}{10}\right)^{-3/2}M_\odot.
\end{equation}

In the following sections we examine how $f_*$ depends on galaxy mass, which is important for calculating the UV luminosity function. However, for reionization the relevant quantity is the star formation efficiency $f_*$ integrated over all galaxies at a given time, so in this section we do not consider the mass dependence of $f_*$.

Finally, for the redshift dependence of the clumping factor we use the fitting formula obtained from numerical simulations for the standard $\Lambda$CDM model \cite{Meletal06}:
%(9)
\begin{equation}
C(z)=27.466\exp\left(-0.114z+0.001328z^2\right).
\label{cfac}
\end{equation}
In our case, clumping is larger at higher halo masses but suppressed at small masses due to the power spectrum cutoff. Since numerical simulations for a non-standard spectrum with a cutoff have not previously been performed, we use expression~(\ref{cfac}) as an estimate; this is permissible because the clumping factor varies slowly with redshift and is of order $C\sim10$.

%============================================================ Fig 3
\begin{figure}
\centering
\includegraphics[width=0.48\textwidth]{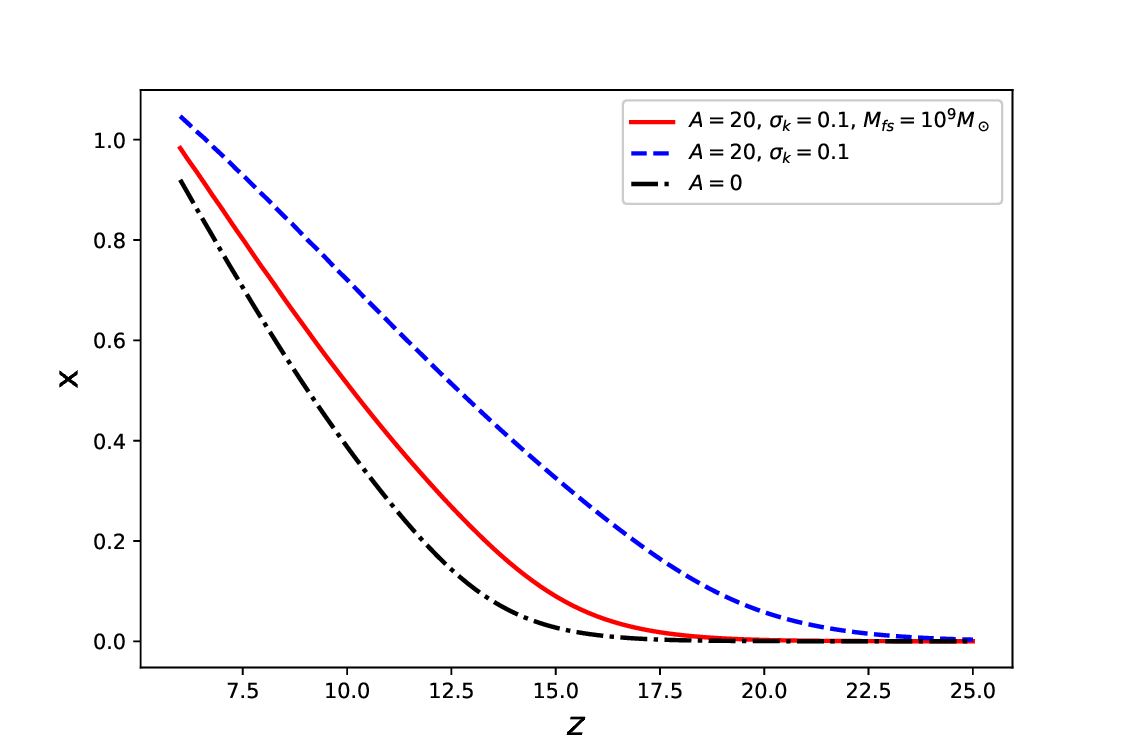}
\caption{Ionization fraction $x$ as a function of redshift $z$. The black dot-dashed line corresponds to the standard $\Lambda$CDM model, the blue dashed line to the bump model, and the red solid line to the model with a bump and a cutoff.}
\label{gr3}
\end{figure}
%=================================================================

The solution of Eq.~(\ref{ioneq}) is shown in Fig.~\ref{gr3}. As seen from the figure, in the bump model the epoch of 50\% reionization shifts to $z\sim15$, which is inconsistent with $z_{re} = 7.68\pm 0.79$ \cite{zref} obtained from the Planck CMB anisotropy analysis. The model with both a bump and a cutoff yields approximately the same reionization history as the standard $\Lambda$CDM model and is broadly consistent with observations: 50\% ionization in this model is reached at $z\simeq10$, close to the observational estimate.

%%%%%%%%%%%%%%%%%%%%%%%%%%%%%%%%%%%%%%%%%%%%%%%%%%%%%%%%%%
\section{Galaxy merger rate}
%%%%%%%%%%%%%%%%%%%%%%%%%%%%%%%%%%%%%%%%%%%%%%%%%%%%%%%%%%

The early galaxies, a significant fraction of which may have formed via the monolithic mechanism, subsequently merge with other galaxies and build up their mass. Thus, in later generations of galaxies the hierarchical merger picture is restored. The infall of a small galaxy into a larger one can be accompanied by a starburst, and at high redshift such galaxies undergoing localized star formation episodes are likely what we observe. In this section we examine the statistics of galaxy mass growth events; in the following section we apply this theory to compute the UV luminosity function.

At first, we need the statistical distribution of mergers in which a halo of a given mass is accreted by a larger halo. As will be shown below, it is precisely this quantity that determines the UV luminosity of a galaxy during a merger-triggered starburst. Previous numerical simulations were limited to standard models without bumps, and although distributions over accreted halo masses were obtained \cite{Steetal08, Genetal09, TraCenMan15}, those results cannot be applied to power spectra with a bump. Various aspects of hierarchical DM halo merging, including halo growth rate calculations, have been considered within the EPS framework in numerous works, see e.g. \cite{LacCol93, SomKol99, NeiBosDek06, NeiDek08, MorGioShe08, ParColHel08, Nadetal23}. However, we have not found in the literature the statistical distribution in the form described above, and therefore in this work we present its derivation within EPS, supplemented by physical requirements on the merger process. The resulting statistical distribution shows agreement with new $N$-body simulations we performed for models with bumps, which justifies applying it to more complex spectra combining a bump and a small-scale cutoff.

Within EPS, by analyzing an ensemble of density perturbation ($\delta$) trajectories, one can compute the probability density for crossing the moving barrier $\omega\equiv\delta_c/D(z)$ \cite{LacCol93}. To this end, let us consider the ensemble of galaxies (virialized halos of mass $M_1$) at time $t_1$ that by a later time $t_2>t_1$ have been incorporated into more massive halos of mass $M_2$. The probability of crossing the barrier $\omega_1=\omega(z_1)$ in the interval $dS_1$ is \cite{LacCol93}
%%(10)
\begin{equation}
f_1dS_1=\frac{\omega_1}{(2\pi)^{1/2}S_1^{3/2}}\exp\left[-\frac{\omega_1^2}{2S_1}\right]dS_1,
\label{f1ds1}
\end{equation}
where $S(M)\equiv \sigma^2(M)$ (see Eq.~(\ref{dispdef})), and the conditional probability distribution for descendant halos (\cite{LacCol93}, Eq.~(2.16)) takes the form
%(11)
\begin{eqnarray}
f_2(S_2,\omega_2 &|& S_1,\omega_1)dS_2=
\label{f2ds2}
\\
\nonumber
=&&\frac{\omega_2(\omega_1-\omega_2)}{(2\pi)^{1/2}\omega_1}
\left[\frac{S_1}{S_2(S_1-S_2)}\right]^{3/2}
\\
\times&&
\exp\left[-\frac{(\omega_2S_1-\omega_1S_2)^2}{2S_1S_2(S_1-S_2)}\right]dS_2.
\nonumber
\end{eqnarray}

At this stage of the calculation a paradox arises. For small time interval $\Delta t\to 0$, formally infinite accretion rate is possible, occurring when a halo of finite (but not infinitesimally small) mass is absorbed by a larger halo. This unphysical property is related to the EPS approximation, in which halo formation is treated as an instantaneous process occurring at the moment the density trajectory crosses the barrier $\omega$. We can overcome this issue by imposing the physical requirement that halo formation and the merging of two halos is not instantaneous but takes approximately one dynamical time, 
%(12)
\begin{equation}
t_d=\frac{\gamma(z)}{\left(6\pi G \rho_h\right)^{1/2}},
\label{tdeq}
\end{equation}
where the halo density is $\rho_h(z)=\varkappa \Omega_m\rho_{cr}(1+z)^3$,
$\varkappa=18\pi^2$, and the function $\gamma(z)\sim1$ will be determined below. Approximately this time interval is required for the small accreting halo to fall into the central region of the larger halo before star formation begins. It is therefore reasonable to assume $\dot M=M_1/t_d(z)$, giving numerically
%(13)
\begin{equation}
\dot M=36\left(\frac{M_1}{10^9M_\odot}\right)\left(\frac{1+z}{13}\right)^{3/2}\gamma^{-1}(z)\frac{M_\odot}{\mbox{yr}}.
\label{dotm1}
\end{equation}

The underlying physical processes can be described as follows. A small virialized halo accreting onto a more massive halo already contains trapped gas. Most of this gas may not be spread throughout the halo but concentrated in a compact disk at the halo center. Within one dynamical time, the small halo falls toward the central region of the galaxy (the large halo), which may also contain a dense gas disk, although some gas may remain in the extended halo, analogous to a gas halo of the Galaxy. The small halo continues to hold its gas as it is transported to the center of the large halo. In the central regions, where the gas density is high, the gaseous component is decelerated. The small DM halo itself flies on further (analogous to the Bullet Cluster), while in the gas cloud undergoes compressions and shock waves, which trigger efficient star formation in the central region of the large halo

As the small halo falls toward the central region, it is subject to tidal stripping by gravitational forces and may lose some mass at the periphery.
However, if the mass of the small halo is roughly an order of magnitude less than the mass of the large halo, most of the small halo survives and reaches the stellar disk region of the large halo. Another process that could impede the transport of gas trapped in the small halo is ram pressure from the gas halo of the large galaxy. If this force were sufficiently large, it could strip the gas from the small halo, while the DM halo itself (now devoid of gas) would continue on its trajectory.

Let us estimate the galaxy parameters for which this effect becomes significant. 
If $\rho_g$ is the gas density in the large halo at some current distance from the center, and $v$ is the velocity of the small halo (of order the virial velocity of the large halo), then the gravitational force retaining all the gas in the small halo is
%(14)
\begin{equation}
F_{\rm grav}\sim\frac{GM_1^2f_b}{R_1^2},
\end{equation}
where $f_b=\Omega_b/\Omega_{DM}$ is the mass fraction of baryons, $\Omega_b$ and $\Omega_{DM}$ are the cosmological baryon and DM density parameters, respectively. Let the characteristic disk size in the small halo be a fraction $\lambda$ of the virial radius, then the ram pressure force $F_{\rm hydro}$ acting on the disk of the small halo is then of order the product of the ram pressure $p=\rho_gv^2/2$, and the cross-section area is about $\pi R_1^2\lambda^2$:
%(15)
\begin{equation}
F_{\rm hydro}\sim\frac{1}{2}\rho_gv^2\pi R_1^2\lambda^2.
\end{equation}
From the condition $F_{\rm grav}>F_{\rm hydro}$ we obtain the gas retention condition for the small halo, which can be expressed numerically as
%(16)
\begin{eqnarray}
1+z&>&1.9\left(\frac{\rho_g}{10^{-27}\mbox{~g~cm$^{-3}$}}\right)^{1/4}\left(\frac{\lambda}{0.1}\right)^{1/2}
\label{odinz}
\\
&\times&\left(\frac{v}{200\mbox{~km~s$^{-1}$}}\right)^{1/2}\left(\frac{M_1}{10^9M_\odot}\right)^{-1/6},
\nonumber
\end{eqnarray}
where Milky Way parameters have been substituted for the large halo in the estimates. Thus, gas can be retained in the small halo for galaxies at redshifts $z\geq1$, and the merger-driven starburst mechanism described above can operate at high $z$. The delivery of the baryonic disk of the small halo to the central region of the large halo proceeds without obstruction. The outcome, however, depends on the residual gas density $\rho_g$ in the large halo, which in high-$z$ galaxies may differ from that in the Milky Way.

One could also consider an alternative approach in which the mass growth is attributed to the total DM flow onto the galaxy both in the form of virialized halos and as a diffuse matter. However, when the diffuse component (a mixture of DM and baryons) is accreted, the gas remains on the periphery of the large halo and moreover has low density. By contrast, a small virialized halo rapidly transports gas to the central region within a dynamical time, triggering a starburst as described above. Thus, it is not the total mass growth of the large halo (including the diffuse component) that drives star formation, but only the part associated with accretion of already virialized halos, although in some cases the diffuse accretion may also influence star formation at later stages. It should be noted that the works \cite{Deketal09, StiRic26} considered models in which star formation bursts in early galaxies were not associated with mergers, but occurred repeatedly due to internal processes.

The interval $\Delta\dot M$ receives contributions from pairs with different masses $M_1$ and $M_2$ under the condition that the resulting accretion rate $\dot M$ is the same. We need to select, from halos of mass $M_1$ in the interval $dS_1$, those that within the time interval $t_d$ will be incorporated into halos of mass $dM_2$, determine their number density, express $dS_1$ in terms of $\dot M$, and then integrate over all $M_2>2M_1$. The factor of ``2'' arises because the final mass $M_2$ includes both $M_1$ and the mass of the progenitor halo with which $M_1$ merged; restricting to accretion of small halos onto more massive ones requires $M_2>2M_1$. Otherwise, the roles of $M_1$ and the progenitor of $M_2$ would be  interchanged. In other words, when we observe the merger of two halos of different masses, we define the smaller halo as being accreted by the larger one, and not vice versa; consequently, the mass growth event of the smaller halo in such a process should not be counted in the overall statistics. In this framework, the distribution of the rate at which small halos are incorporated into larger ones (number of such events per Mpc$^3$ per logarithmic mass interval) is
%(17)
\begin{equation}
\frac{dn}{d\ln\dot M}=\int\limits_{M_{\rm min}}^{M_{\rm max}}dM_2\frac{\rho_{DM}}{M_1}
f_2\frac{dS_2}{dM_2}f_1\frac{dS_1}{dM_1}\frac{dM_1}{d\ln\dot M},
\label{dndmuv}
\end{equation}
where $M_{\rm min}=2M_1$, $M_{\rm max}\sim10^{12}M_\odot$, $\dot M=M_1/t_d(z)$, and the last derivative is equal to $M_1$. The result changes little when $M_{\rm max}$ is increased above $10^{12}M_\odot$, since such halos are very rare at $z>6$. Since $t_d=t/\varkappa^{1/2}\ll t$, one can approximate $\omega_1-\omega_2 =(d\omega/dt)t_d = 2\omega/(3\varkappa^{1/2})$.

The function $\gamma(z)$ entering equation~(\ref{tdeq}) is difficult to estimate theoretically, so we determine it by requiring the best agreement between formula~(\ref{dndmuv}) and the numerical simulations described below. For $z=7$, 12, and 17 we obtain $\gamma=1$, 1.1, and 1.5, respectively. As a fitting function over the interval $z=7$--$17$ we choose
%(18)
\begin{equation}
\gamma(z)=1.095-0.017z-0.00078z^2+0.00019z^3.
\label{ganf}
\end{equation}

The growth of $\gamma(z)$ with increasing $z$ has a simple explanation. Two merging halos are still distinct when the separation between their centers exceeds the sum of their virial radii. For a given $\dot M$, at lower redshift this value is realized predominantly by the infall of a much less massive halo onto a large one. Since the small halo is much smaller than the large halo, the infall begins at approximately the virial radius of the large halo and lasts roughly one dynamical time. By contrast, for the same $\dot M$ at higher redshift, halos of comparable sizes predominantly merge (since large halos are rare at that time), so at the onset of the merger the center of the smaller halo is already at some distance beyond the virial radius of the larger one. Accordingly, the center of the small halo requires not only one dynamical time of the larger halo to reach its center, but also additional travel time to first arrive at the virial radius of the larger halo.

Three numerical simulations were used to verify the analytical calculations. All three have a box size of 47~Mpc, $1024^3$ DM particles, and start from $z_{init}=120$; initial conditions were generated with the \texttt{Ginnungagap}\footnote{https://github.com/ginnungagapgroup/ginnungagap} \cite{Pilipenko26}, and the simulations themselves were run with \texttt{GADGET-2} \cite{Springel05}. The simulations differ only in their initial power spectra: the first uses the standard spectrum $P_{\Lambda CDM}$, while the other two use the bump spectrum $P_\mathrm{bump}$ (\ref{bump}) with $k_0=4.69$ and $k_0=9.38$. Halos were identified with the \texttt{Rockstar} code \cite{Behroozi13a}, and mass accretion histories were analyzed from merger trees constructed with \texttt{ConsistentTrees} \cite{Behroozi13b}.

\begin{figure}[!ht]
\centering
\includegraphics[width=0.48\textwidth]{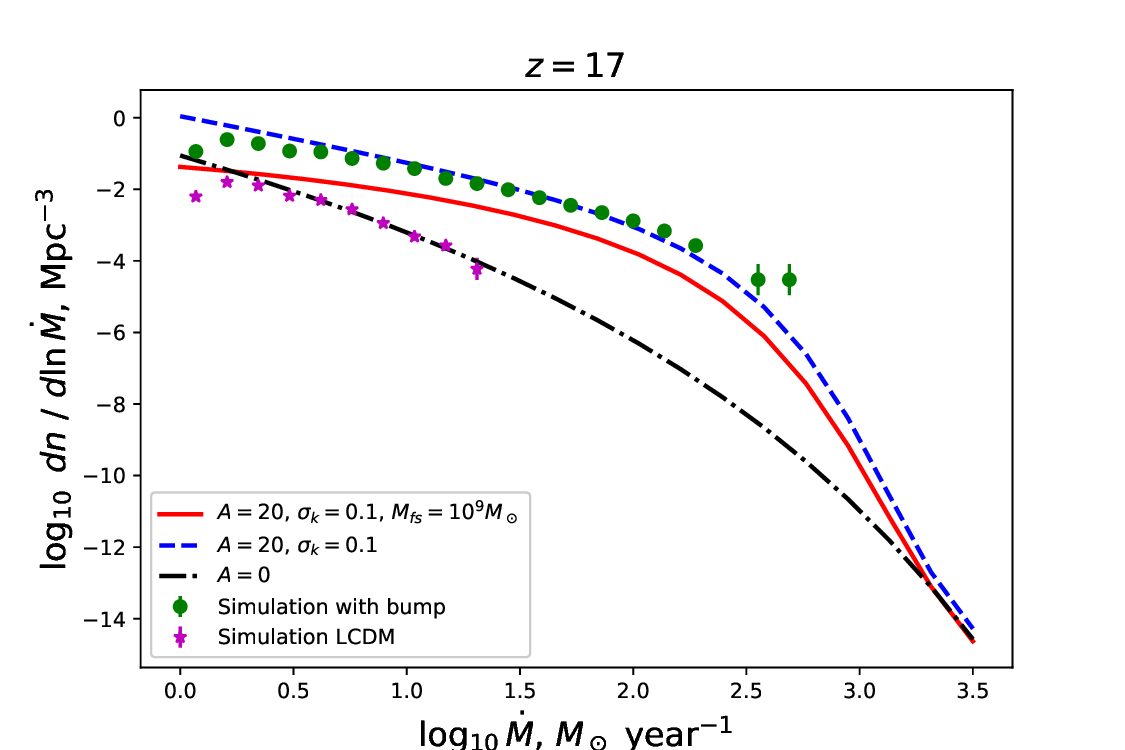}\\
\includegraphics[width=0.48\textwidth]{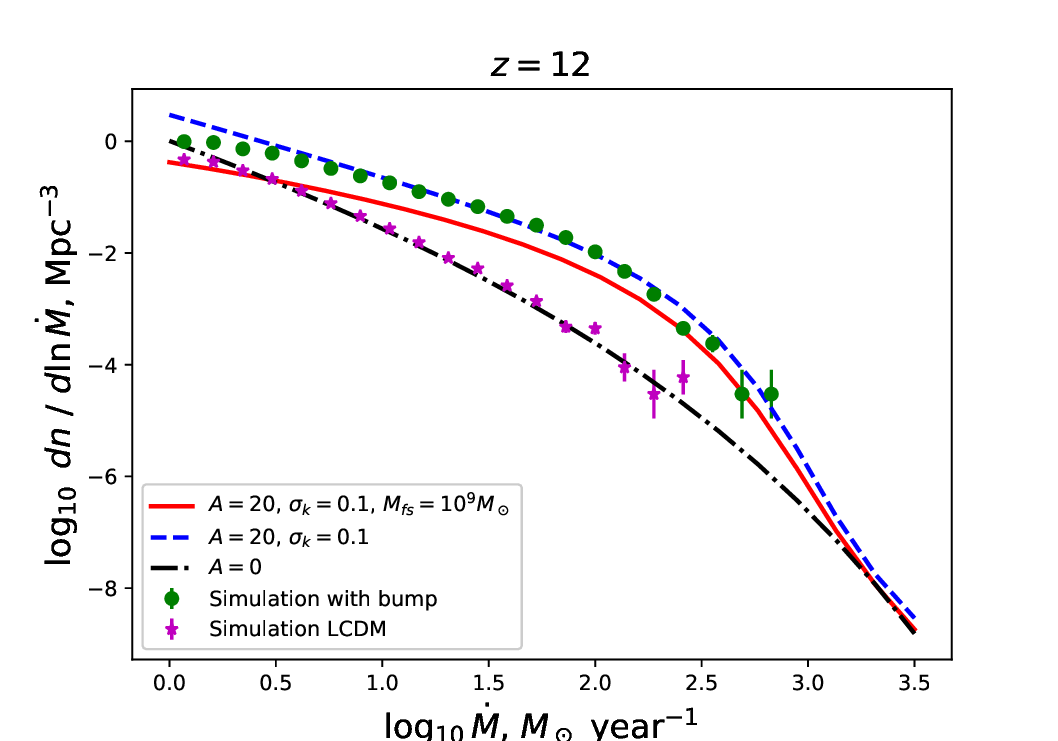}\\
\includegraphics[width=0.48\textwidth]{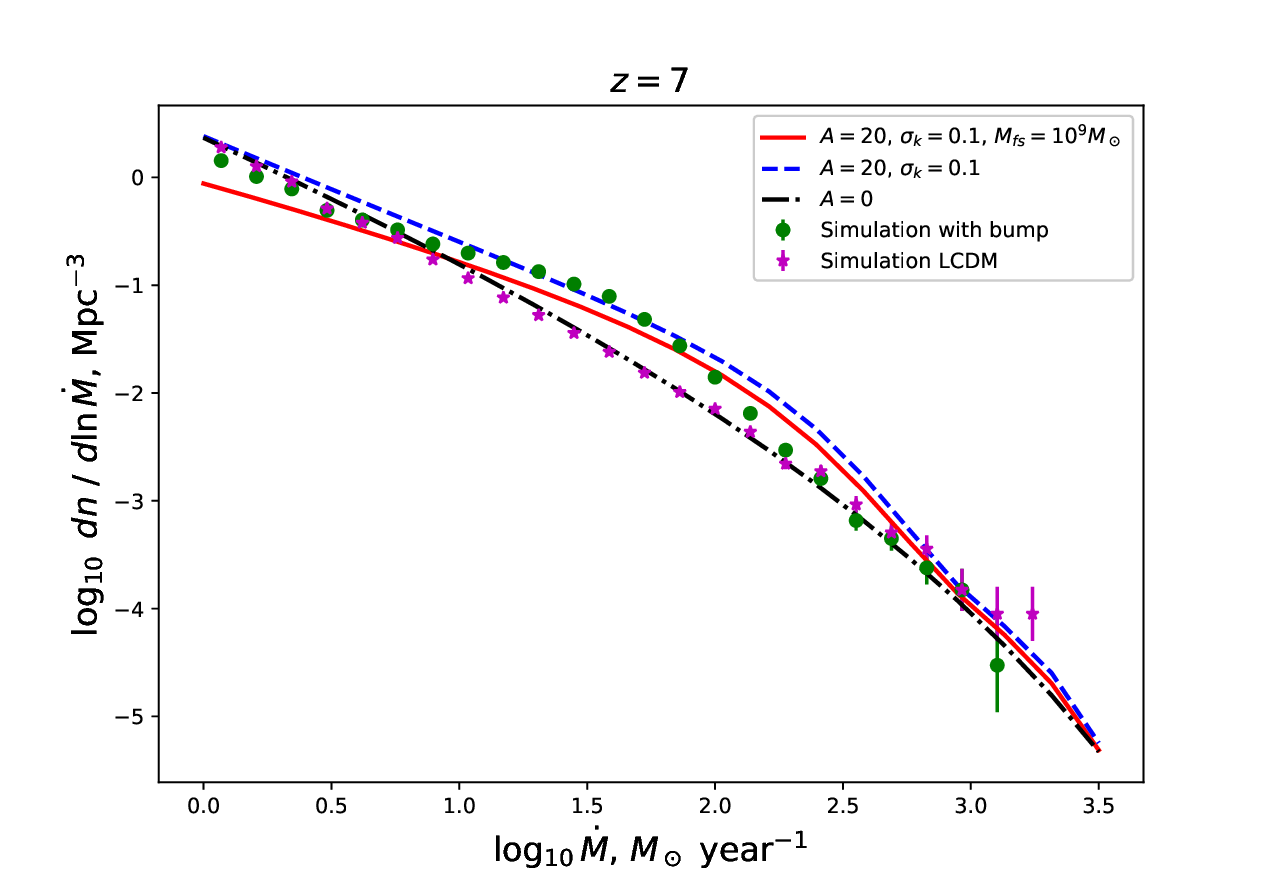}
\caption{Rate at which small virialized DM halos are incorporated into larger halos at $z=17$ (top), $z=12$ (middle), and $z=7$ (bottom), computed from Eqs.~(\ref{dndmuv}) and~(\ref{ganf}) for different perturbation spectra. Points show the $N$-body simulation data.}
\label{grrate}
\end{figure}

\begin{figure}[!ht]
\centering
\includegraphics[width=0.48\textwidth]{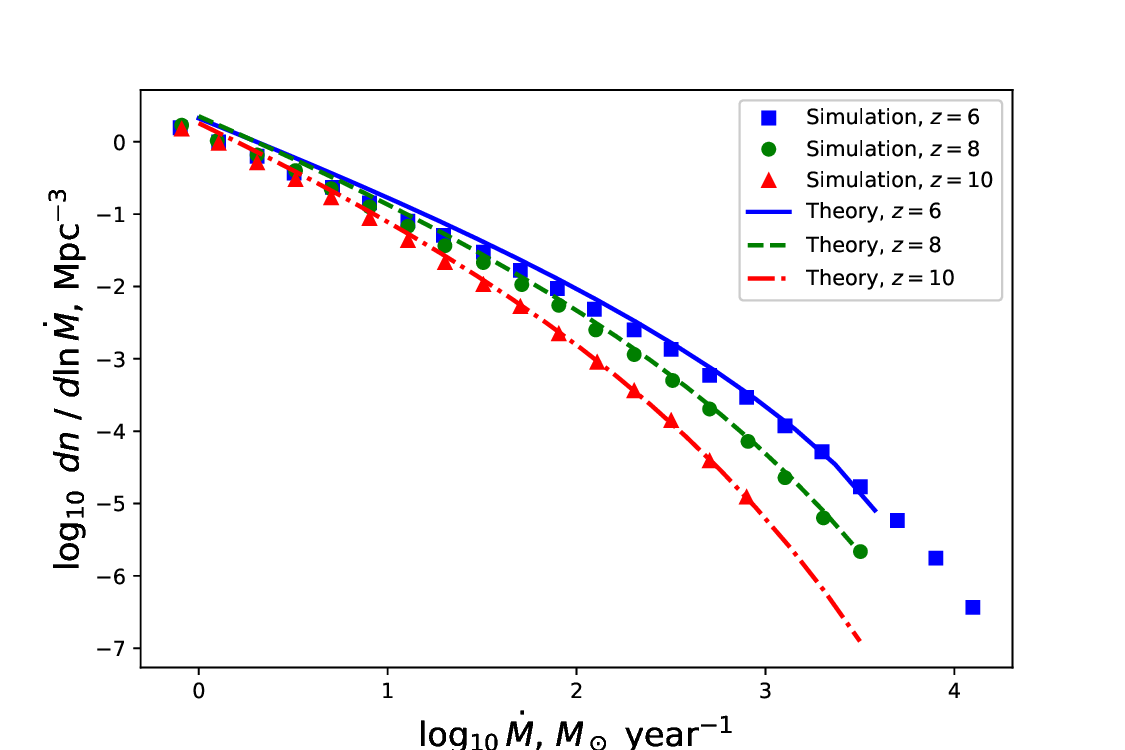}
\caption{Rate at which small virialized DM halos are incorporated into larger halos, computed from Eqs.~(\ref{dndmuv}) and~(\ref{ganf}), compared with the simulation data of \cite{TraCenMan15} for $z=6$, 8, and 10 (the vertical axis shows decimal logarithms). Note that \cite{TraCenMan15} reports the distribution per decade in accretion rate, $dn/d\log_{10}$.}
\label{grtrac}
\end{figure}

The results of formula~(\ref{dndmuv}) are shown in Fig.~\ref{grrate} for redshifts $z=17$, 12, and 7, compared with our $N$-body simulation results.
Simulations are available only for the standard power spectrum and the bump spectrum (Fig.~\ref{grrate} shows only the $\Lambda$CDM and the bump with $k_0=4.69$; results for the third simulation are omitted to avoid cluttering the figure, but they show good agreement). Running a simulation for the spectrum with both a bump and a cutoff is difficult due to accumulating numerical errors in the absence of small-scale perturbations. However, if agreement is found for the first two cases, one can expect EPS to describe
the case with a cutoff equally well. Good agreement between the analytical calculation and the simulations is evident, except at the lowest accretion rates where the simulations have limited resolution. Figure~\ref{grtrac} shows an analogous comparison with the simulation data of \cite{TraCenMan15} for the standard spectrum (without a bump).

In the final calculation one could replace (\ref{f1ds1}) with the Sheth--Tormen function, which accounts for halo ellipticity:
%(19)
\begin{equation}
f_1=A_{ST}a^{1/2}\left(1+\left(\frac{S_1}{a\omega_1^2}\right)^p\right)\frac{\omega_1}{(2\pi)^{1/2}S_1^{3/2}}\exp\left[-\frac{a\omega_1^2}{2S_1}\right].
\end{equation}
However, with the standard parameter choices $a=0.707$, $A_{ST}=0.3222$, and $p=0.3$, the discrepancy with the simulation results remains at the $\sim30\%$ level for different $\gamma(z)$; agreement improves noticeably only when $a=1$ is adopted. Therefore, in this work we use the EPS variant with functions~(\ref{f1ds1}) and~(\ref{f2ds2}), corresponding to spherical collapse. It is precisely in this variant of the theory that the conditional probabilities~(\ref{f1ds1}) and~(\ref{f2ds2}) were derived, and using the Sheth--Tormen function in this context would be logically inconsistent.

%===================================================================fig 4
%====================================================================
%===================================================================fig 5
%====================================================================

Thus, in this section we have developed an efficient algorithm for computing the statistical distribution of halos by their growth rate, showing good agreement with numerical simulation results including for the bump spectrum. Specifically, the distribution is computed from formula~(\ref{dndmuv}) using the fitting function~(\ref{ganf}). This growth-rate distribution provides access to the galaxy UV luminosity distribution, which is the subject of the next section.

%%%%%%%%%%%%%%%%%%%%%%%%%%%%%%%%%%%
\section{Evolution of the galaxy \\ UV luminosity function}
%%%%%%%%%%%%%%%%%%%%%%%%%%%%%%%%%%%

With the launch of the JWST, which observes in the infrared, the rest-frame UV emission of galaxies at redshifts $z\geq10$ became accessible, while the optical emission of these galaxies is cosmologically redshifted to longer wavelengths that are not yet easily observed. The UV emission of a galaxy is dominated by young massive stars, so the UV luminosity function traces the star formation rate. In this section we compute the UV luminosity function associated with starburst episodes triggered by galaxy mergers.

The basic ideas for computing the galaxy UV luminosity function were discussed, for example, in \cite{FerPalDay23, Blaetal25, Peretal25, Aaretal25}, which also provide the relevant observational data and references. In \cite{Zhuetal26} the UV luminosity function in a bump model was computed for the first time, but the galaxy merger rate (formula~(10) in \cite{Zhuetal26}) was taken from the Millennium simulation for the standard $\Lambda$CDM model without a bump. In \cite{Blaetal25} the influence of cosmic string loops on the formation of the first galaxies and constraints from the observed UV signal were studied. In the present work we do not consider topological defects and account only for adiabatic density perturbations arising from a non-standard perturbation spectrum. Moreover, in contrast to other works, we apply the merger-driven accretion rate derived in the previous section (the rate at which smaller virialized halos are accreted by larger halos) taking into account the non-standard density perturbation spectrum.

The mean UV luminosity of a galaxy can be expressed as \cite{Blaetal25}
%(20)
\begin{equation}
L_{UV}=\frac{\dot M_*}{\varkappa_{UV}},
\label{uveq}
\end{equation}
where $\dot M_*$ is the star formation rate in the region of interest, and $\varkappa_{UV}=1.15\times10^{-28}M_\odot\,\mbox{yr}^{-1}/(\mbox{erg}\,\mbox{s}^{-1})$.
The relation between UV luminosity and absolute UV magnitude is
%(21)
\begin{equation}
\log_{10} L_{UV}=0.4\left(51.6-M_{UV}\right).
\label{dotmz}
\end{equation}

The star formation rate $\dot M_*$ is related to the DM halo mass accretion rate $\dot M$ by
%(22)
\begin{equation}
\dot M_*=f_*(M)f_b\dot M.
\label{dotmzvmh}
\end{equation}
In this approach only the freshly accreted gas is considered, and star formation from pre-existing gas is neglected. This is justified by the more vigorous star formation driven by galaxy mergers compared to the slow evolution of the interstellar medium due to internal processes alone. To describe the star formation efficiency during accretion onto a halo of mass $M$ we use \cite{MirFurSun17}
%(23)
\begin{equation}
f_*(M)=\frac{2\varepsilon_*}{(M/M_c)^{-\alpha_*}+(M/M_c)^{-\beta_*}},
\label{starfor}
\end{equation}
where $\varepsilon_*$ is the normalization, $\alpha_*>0$, $\beta_*<0$, and $M_c$ is the characteristic mass. Expression~(\ref{starfor}) is consistent with both simulation results and observational data, and holds across various star formation scenarios.

It should be noted that when the gas disk of the small halo impacts the gas disk of the large halo, the gas in the latter also participates in star formation, and these two gas masses are comparable. Therefore, in the model considered here, the fraction of gas converted into stars is effectively computed from twice the mass of the accreted gas. The shape of function~(\ref{starfor}) is motivated by two theoretical arguments: at the high-mass end, the downturn is associated with the formation of central supermassive
black holes, whose radiation suppresses star formation in the galaxy; 
at the low-mass end, the expanding shells of the first supernovae expel gas from the halo, suppressing subsequent star formation. That is, starbursts may occur in low-mass halos but are short-lived. 
The parameters in~(\ref{starfor}) may have differed at different epochs and cannot yet be fixed by theory, leaving considerable freedom. Ultimately, these parameters are determined from the best fit to observations, as was done, for example, in \cite{Aaretal25}. This parametric freedom introduces uncertainty into the final result, and at the present stage one can only state that there exist parameter sets for which theory and observations agree.

%=================================== table 1
\begin{table}[tbp]
\centering
\caption{\label{tab} Fitting parameters in formulas~(\ref{starfor}) and~(\ref{dndmuv2})
for different redshifts.}
\begin{tabular}{|c|c|c|c|c|c|}
\hline
$z$ & $\varepsilon_*$ & $M_c/M_\odot$ & $\alpha_*$ & $\beta_*$ & $f_{\rm eff}$ \\
\hline
17 & 0.8 & $5\times 10^{11}$ & 0.5 & -0.5 & 0.02 \\
\hline
12 & 0.2 & $5\times 10^{11}$ & 0.5 & -0.5 & 0.03 \\
\hline
7 & 0.1 & $5\times 10^{11}$ & 0.75 & -0.95 & 0.06 \\
\hline
\end{tabular}
\end{table}
%======================================================

For a given mass $M$ and the corresponding $f_*=f_*(M)$, the accreted small halo mass and the mass accretion rate are expressed in terms of the galaxy
absolute UV magnitude as
%(24)
\begin{equation}
M_1=3\times10^9e^{-0.92(M_{UV}+19)}\left(\frac{1+z}{13}\right)^{-3/2}\!\!\!\!\!\!\!\gamma(z)\left(\frac{f_*}{0.1}\right)^{-1}\!\!\! \!\!M_\odot,
\label{m1msmvu}
\end{equation}
%(25)
\begin{equation}
\dot M=1.2\times10^2e^{-0.92(M_{UV}+19)}\left(\frac{f_*}{0.1}\right)^{-1}\frac{M_\odot}{\mbox{yr}}.
\label{m1msmvu2}
\end{equation}
These two quantities are related by Eq.~(\ref{dotm1}).

The galaxy halo mass growth rate $\dot M$ is often estimated as exponential, $\dot M_h\propto \exp(-a_*z)$ \cite{Wecetal02, Deketal13}, where the constant $a_*$ is determined from large simulations since theory does not fix its value (see \cite{NeiBosDek06}). In the present work we cannot use this recipe, since for our non-standard spectrum the hierarchical clustering pattern and the value of $\dot M$ may be substantially different. For this reason we derived $\dot M$ from EPS in the preceding section.

It must be taken into account that by no means every merger will result in the gas disk of the small halo falling onto the gas disk at the center of the large halo. The surrounding density perturbations create tidal forces that impart angular momentum to all small halos falling onto the large one. As a result, in most cases a small halo will follow a wide (non-radial) orbit. On such an orbit, the gas disk of the small halo will not collide with the disk of the large halo. Only gradually, through dynamical friction, will the small halo approach the central region and eventually reach the center (for large enough small halo mass), but by that time the gas disk of the small halo may have already dissolved into the volume of the large halo. A detailed study of these processes is possible only with high-resolution, long-duration numerical simulations. Therefore, here we provide only an estimate of the fraction $f_{\rm eff}$ of small halos that fall directly to the center of the large halo within one dynamical time. For this purpose we use the result of \cite{BerDokEro03} for the pericenter distance $R_c$ of an accreting particle from the center of the large halo relative to its virial radius $R_v$:
%(26)
\begin{equation}
\frac{R_c}{R_v}\simeq0.3\nu^{-2}f^2(\delta_{\rm eq}),
\label{rcrv}
\end{equation}
where $\nu=\delta/\sigma(M)$ is the peak height in the density perturbation field (at the matter-dominated onset $\delta(t_{\rm eq})=\delta_{\rm eq}$) from which the large halo formed, and $f(\delta_{\rm eq})\simeq1.2$ in the mass range of interest. Setting the disk size to $\sim0.1\,R_v$, a disk collision requires $R_c\sim 0.1R_v$. This condition is satisfied for peaks with $\nu\geq2$. The fraction of such peaks in a Gaussian distribution over $\nu$ (counting only positive perturbations) is $f_{\rm eff}\sim0.025$.
This estimate gives only the order of magnitude; below we treat $f_{\rm eff}$ as a free parameter to be fixed by observational data.

We note that in previous UV luminosity function calculations based on models without a bump, the coefficient $f_{\rm eff}$ was implicitly set to unity, i.e., the small probability of a central collision within one dynamical time was not accounted for. Accounting for this small probability would lead to a severe deficit in the number of merger-driven starburst events. In the bump model considered here, where the merger rate is significantly enhanced, including $f_{\rm eff}$ leads to a self-consistent picture consistent with the observations.

The galaxy UV luminosity function can be computed by a formula analogous to~(\ref{dndmuv}), with minor modifications and including the fraction $f_{\rm eff}$. The interval $\Delta M_{UV}$ receives contributions from galaxies with different masses $M_1$ and $M_2$ under the condition that the resulting star formation rate $\dot M_*$ is the same. In this framework the UV luminosity function (galaxies per Mpc$^3$ per magnitude) is
%(27)
\begin{equation}
\frac{dn}{dM_{UV}}=f_{\rm eff}\int\limits_{M_{\rm min}}^{M_{\rm max}}dM_2\frac{\rho_{DM}}{M_1}
f_2\frac{dS_2}{dM_2}f_1\frac{dS_1}{dM_1}\frac{d(M_1/t_d)}{dM_{UV}}t_d,
\label{dndmuv2}
\end{equation}
where $M_{\rm min}={\rm max}\{M_L/2,\,M_{\rm cooling}(z)\}$;
the quantity $M_L$ is determined from Eqs.~(\ref{uveq})--(\ref{starfor}) for a given $M_{UV}$ and the condition $\dot M=(M/2)/t_d$ (the origin of the factor of~``2'' was explained before Eq.~(\ref{dndmuv})); 
the remaining quantities are the same as in~(\ref{dndmuv}).
In the literature, an empirical Gaussian probability distribution in $M_{UV}$ is often assumed. We do not adopt this assumption but instead compute the distribution over $M_{UV}$ directly.

\begin{figure}
\centering
\includegraphics[width=0.48\textwidth]{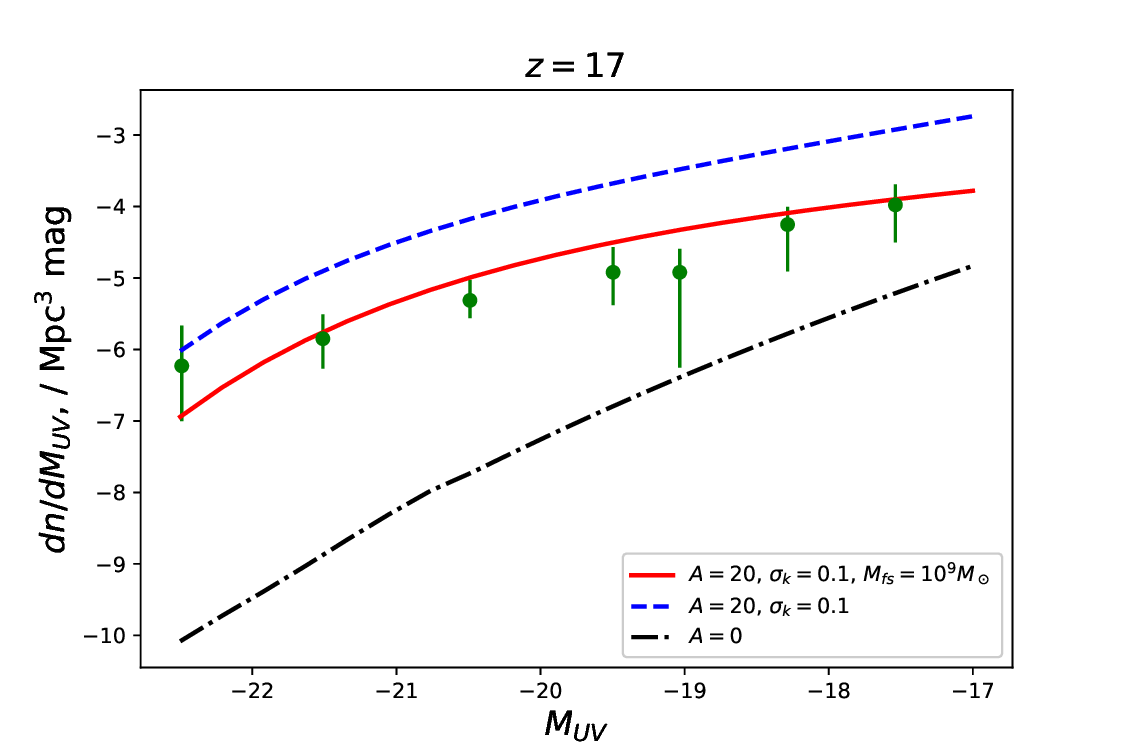}\\
\includegraphics[width=0.48\textwidth]{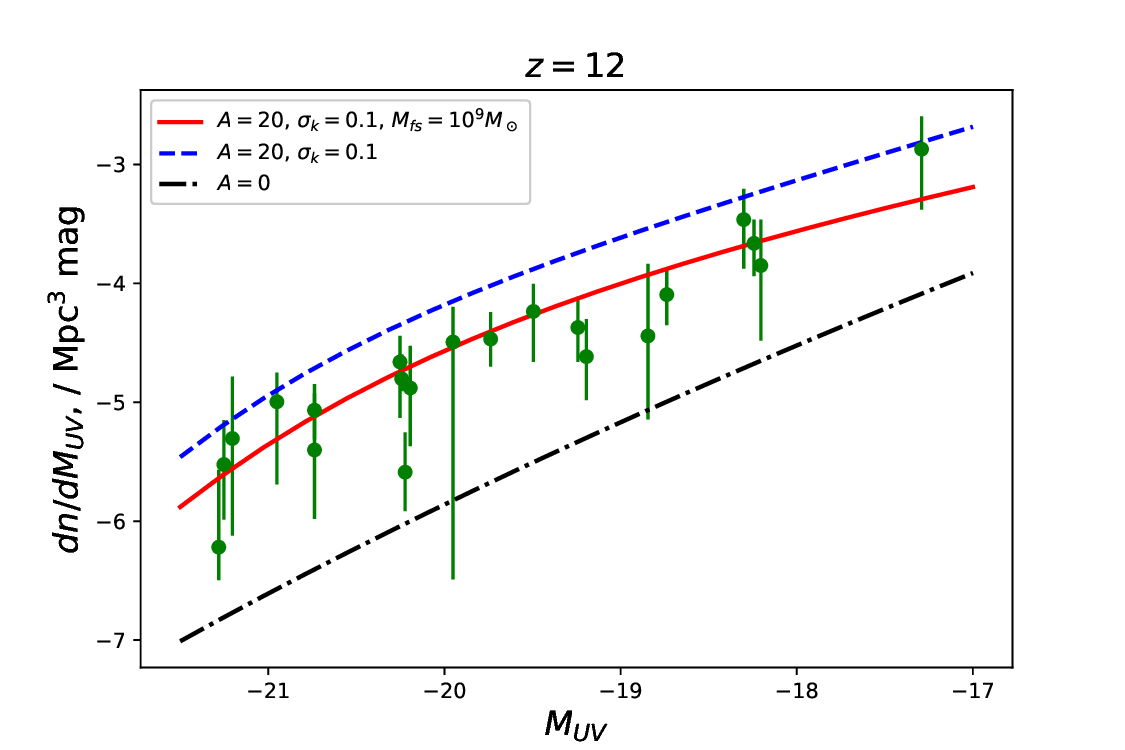}\\
\includegraphics[width=0.48\textwidth]{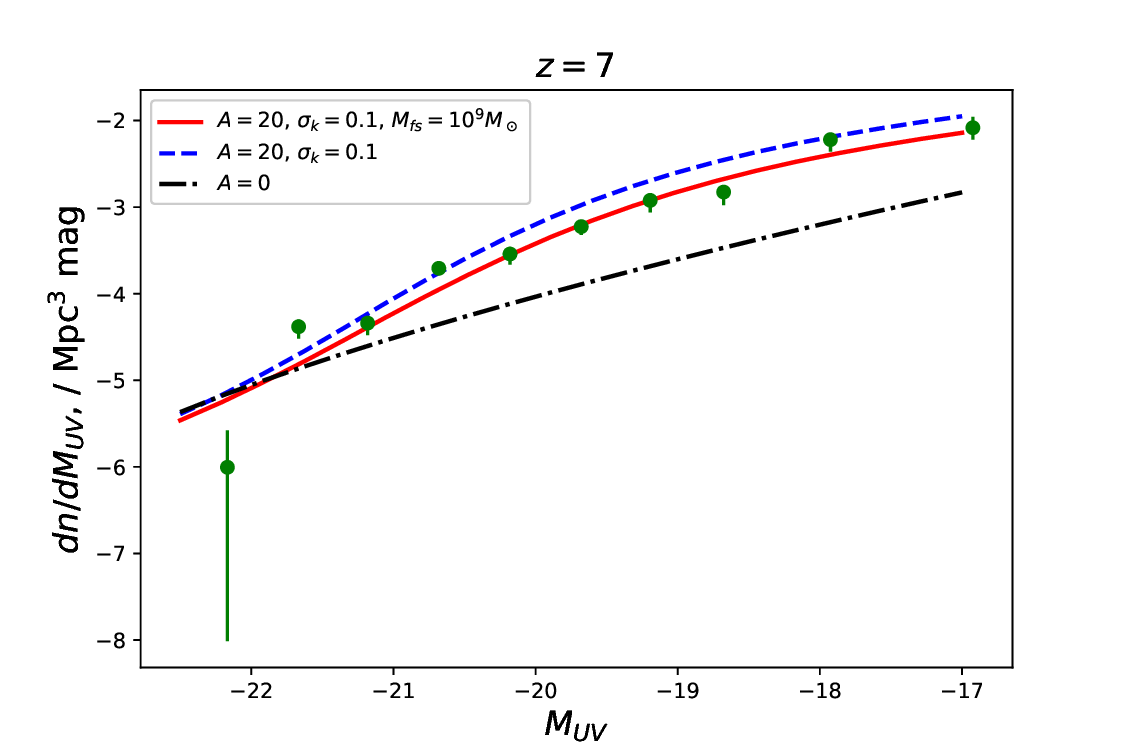}
\caption{UV luminosity function of galaxies at $z=17$ (top), $z=12$ (middle), and $z=7$ (bottom), computed from formula~(\ref{dndmuv2}) for different perturbation spectra. Points show the observational data (the vertical axis shows decimal logarithms).}
\label{grfz}
\end{figure}

The results are shown in Fig.~\ref{grfz} for redshifts $z=17$, 12, and 7, using the parameter sets given in Table~\ref{tab} (analogous parameter sets for models without a bump, within numerical simulations, can be found in \cite{Aaretal25}). The observation data points in Fig.~\ref{grfz} are taken from the work \cite{Aaretal25}. At $z=12$ the observational data are already fairly reliable and in several cases have spectroscopic confirmation. At higher $z$ observational data also exist but are less certain. The bump models with the parameters adopted here reproduce the observational data reasonably well.
The model with only a bump and no cutoff, as was shown above, does not reproduce the cosmic reionization history. We therefore conclude that the density perturbation spectrum must contain, in addition to the bump, a suppression on the small-mass side. At $z=7$ the data for the brightest galaxies do not match the calculation (the leftmost data point), although the observational uncertainties are large.

%=====================================================fig6
%====================================

In this section we considered only starbursts associated with the accretion of small halos onto larger ones. However, in the monolithic theory of galaxy formation, the very first virialization episode of a galaxy should also be accompanied by star formation. Moreover, the number of stars in this case depends on the total gas mass of the forming galaxy, and one might expect this to produce a high-luminosity episode. Perhaps the brightest of the observed galaxies correspond to the initial collapse of gas via the monolithic mechanism. In the left part of Fig.~\ref{grfz} at $z=17$ and $z=12$, an excess of the brightest galaxies above our calculations is indeed visible. It is possible that at these epochs the galaxies undergoing the monolithic formation stage provided a significant contribution, whereas by $z\sim7$ the monolithic phase had largely concluded.

%%%%%%%%%%%%%%%%%%%%%%%%%%%%
\section{Conclusions}
%%%%%%%%%%%%%%%%%%%%%%%%%%%%

Observational data of recent years, primarily from JWST, have posed a series of challenging problems for the standard cosmological model. They are related to the unexpectedly early formation of massive galaxies (see the review \cite{Sil25}). Taking a conservative approach and not invoking new physics at $z\sim10$, one can explain the observations by means of a new density perturbation spectrum, which is more complex than in the standard $\Lambda$CDM model: it contains a bump at the scale corresponding to mass $\simeq10^{10}M_\odot$, and a cutoff at smaller scales. The origin of the bump can be naturally associated with physical processes during the inflationary stage, for example, within the cascade vacuum relaxation \cite{cascade}, while the cutoff may arise from the perturbation spectrum itself or from free-streaming of DM particles in the case of warm DM.

The aim of this work is to demonstrate the self-consistency of a cosmological model with such a sophisticated spectrum, and its agreement with observational data, including the the reionization epoch.

Here we considered three variants of the density perturbation spectrum, of which two are non-standard: one with a bump, and another combining a bump with a small-scale cutoff. We investigated the quantitative agreement of these spectra with three key observational criteria: the observed excess of massive galaxies at $z>6$, the timing of the cosmic reionization epoch, and the shape of the UV luminosity function of early galaxies. We confirm that the standard perturbation spectrum does not explain the galaxy excess. The bump model can explain this excess but substantially shifts reionization to higher
redshifts. The model incorporating both a bump and a cutoff satisfies all three observational criteria formulated in the Introduction. In addition, the suppression of the spectrum amplitude on small scales in the bump-plus-cutoff model is capable of resolving the satellite overabundance problem present in the standard $\Lambda$CDM model.

The galaxy formation model considered in this work has the following main features:
\begin{itemize}
\item Due to the bump in the perturbation spectrum, at $z\geq9$ the most massive galaxies form 1--2 orders of magnitude more abundantly than in the standard cosmological model.
\item The cutoff at scales smaller than the bump scale suppresses the formation of the first stars, compensating for the bump's effect on early star formation. As a result, this model reproduces the reionization history of the Universe in essentially the same way as the standard model.
Note that in the bump-only model (without a cutoff) reionization would shift to unacceptably high redshifts.
\item In the bump-plus-cutoff model, the fraction of massive galaxies forming not hierarchically but through a single collapse, i.e., via the monolithic mechanism, is increased compared to the standard model. This fraction is not very large, and most galaxies still form hierarchically through the merging of smaller galaxies.
\item In this model, the fraction of stars that form directly in large monolithically formed galaxies,  bypassing the stage of metal enrichment in small protogalaxies, is increased. We call such stars Population~IV stars. Supernova explosions of these stars in massive galaxies enrich the surrounding gas with metals, so that a relatively larger fraction of metals is produced in large galaxies.
\item The hierarchical formation mechanism continues to operate for the majority of galaxies, leading to the merging of small galaxies with each other and with large galaxies, some of which formed monolithically.
The new EPS-based method we developed for computing conditional probability distributions yields a merger-rate distribution in good agreement with direct $N$-body simulations.
\item At the centers of merging galaxies lie stellar and gaseous disks.
During a merger, the disk of the small galaxy, embedded in its DM halo, moves toward the center of the large galaxy, where it can collide with the large galaxy's disk and trigger a starburst. Estimates show that only about 3\% of small galaxy disks will collide with the large galaxy's disk within one dynamical time after their merger; in most cases the disks (or dense central regions) of the small galaxies and their DM halos will remain in the halos of large galaxies for a long time, moving along orbits as satellite galaxies.
\item The picture of galaxy mergers and disk collisions described above, which trigger starbursts, successfully reproduces the UV luminosity function of galaxies obtained from HST and JWST observations. The calculation contains parameters describing the dependence of star formation efficiency on galaxy mass.
\end{itemize}

Many recent works attempt to build detailed and complex models of early galaxy evolution; however, a number of key model parameters, such as the star formation efficiency, the UV photon escape fraction, the influence of active central black holes on star formation, and others, remain uncertain. These processes are treated differently within various approximate frameworks. There remains a need to explain the main observational facts with simple, transparent models, although future refinements are of course also important. In the present work we used a minimum of complex model assumptions and the well-established EPS formalism, verifying its predictions against direct $N$-body simulations. We have shown that the model with a non-standard perturbation spectrum explains the main observational facts. Further development and testing of this model may include developing new $N$-body simulation algorithms for an accurate description of large-scale structure formation in a cosmological model with a complex density perturbation spectrum. Further theoretical investigation of the processes of early galaxy and star formation is of course also important. A necessary new ingredient may turn out to be the role of early supermassive black holes.

%%%%%%%%%%%%%%%%%%%%%%%%
\section*{Acknowledgments}
%%%%%%%%%%%%%%%%%%%%%%%%

The work of V.~N.~Lukash, E.~V.~Mikheeva, S.~V.~Pilipenko, and M.V.~Tkachev was supported within the State assignment of the P.~N.~Lebedev Physical Institute, project No.~FFMR-2024-0013. The work of Yu.N.~Eroshenko was carried out within the state assignment for the Institute for Nuclear Research of the Russian Academy of Sciences.

%%%%%%%%%%%%%%%%%%%%%%%%%%%


\begin{thebibliography}{99}
%%%%%%%%%%%%%%%%%%%%%%%%%%%
%1
\bibitem{BICEPlanck}
P.A.R. Ade, et al., Improved constraints on primordial gravitational waves using Planck, WMAP, and BICEP/Keck observations through the 2018 observing season, Phys. Rev. Lett. \textbf{127}, 151301 (2021).
%2
\bibitem{linde2025} A. Linde, Alexei Starobinsky and modern cosmology, arXiv:2509.01675.
%3
\bibitem{cascade} V.N. Lukash, E.V. Mikheeva, Cascade relaxation of the gravitating vacuum as a generator of the evolving Universe, JETP Lett. \textbf{121}, 421 (2025).
%4
\bibitem{teleparalel} I.V. Fomin, S.V. Chervon, L.K. Duchaniya, B. Mishra, The scalar-torsion gravity corrections in the first-order inflationary models, Physics of the Dark Universe \textbf{48}, 101895 (2025).
%5
\bibitem{SPT} E. Camphuis et al., SPT-3G D1: CMB temperature and polarization power spectra and cosmology from 2019 and 2020 observations of the SPT-3G Main field, arXiv:2506.20707.
%6
\bibitem{ACT} T. Louis et al., The Atacama Cosmology Telescope: DR6 power spectra, likelihoods and $\Lambda$CDM parameters, J. Cosmol. Astropart. Phys. 062 (2025).
%7
\bibitem{Naidu2022b} R.P. Naidu, et al., Two remarkably luminous galaxy candidates at $z\simeq 10-12$ revealed by JWST, Astrophys. J. Lett. \textbf{940}, L14 (2022).
%8
\bibitem{Castellano22} M. Castellano, et al., Early results from GLASS-JWST. III. Galaxy candidates at $z \simeq 9-15$, Astrophys. J. Lett. \textbf{938}, L15 (2022).
%9
\bibitem{Finkelstein22}
S. L. Finkelstein, et al., A long time ago in a galaxy far, far away: a candidate $z \simeq 12$ galaxy in early JWST CEERS imaging, Astrophys. J. Lett. \textbf{940}, L55 (2022).
%10
\bibitem{Donnan23} C.T. Donnan, et al., The evolution of the galaxy UV luminosity function at redshifts $z \simeq 8 - 15$ from deep JWST and ground-based near-infrared imaging, Mon. Not. R. Astron. Soc. \textbf{518}, 6011 (2023).
%11
\bibitem{Labbe23} I. Labb{\'e}, et al., A population of red candidate massive galaxies 600 Myr after the Big Bang, Nature \textbf{616}, 266 (2023).
%12
\bibitem{Li_2025_ApJ_981} J. Li, et al., Tip of the iceberg: overmassive black holes at $4 < z < 7$ found by JWST are not inconsistent with the Local $M_{BH}$ --- $M_\star$ relation, Astrophys. J. \textbf{981}, 19 (2025).
%13
\bibitem{zref} N. Aghanim, et al. (Planck Collaboration), Planck 2018 results. VI. Cosmological parameters, Astron. Astrophys. \textbf{641}, A6 (2020).
%14
\bibitem{Tkaetal23} M.V. Tkachev, S.V. Pilipenko, E.V. Mikheeva, V.N. Lukash, Mon. Not. R. Astron. Soc. \textbf{527}, 1381 (2024).
%15
\bibitem{Eroshenkoetal2024} Yu. N. Eroshenko, V.N. Lukash, E.V. Mikheeva, S.V. Pilipenko, M.V. Tkachev, Properties of central regions of the dark matter halos in the model with a bump in the power spectrum of density perturbations, JETP Lett. \textbf{120}, 83 (2024).
%16
\bibitem{TkachevetalPRD2024}
M.V. Tkachev, S.V. Pilipenko, E.V. Mikheeva, V.N. Lukash, Inner structure of dark matter halos at high $z$ in cosmological models with non-power-law primordial spectra, Phys. Rev. D\textbf{110}, 083530 (2024).
%17
\bibitem{TkachevetalPRD2025}
M.V. Tkachev, S.V. Pilipenko, E.V. Mikheeva, V.N. Lukash, High-z SMBHs in cosmological models with enhanced power spectra, Phys. Rev. D \textbf{112}, 063527 (2025).
%18
\bibitem{Eroshenko2025}
Yu.N. Eroshenko, V.N. Lukash, E.V. Mikheeva, S.V. Pilipenko, and M.V. Tkachev, Absorption in the 21 cm hydrogen line at $z > 10$ as a sensitive tool for the construction of a cosmological model on small scales, Astron. Lett. \textbf{51}, 189 (2025).
%19
\bibitem{Qinetal25} W. Qin, S. Kumar, P. Natarajan, N. Weiner, Not-quite-primordial black holes, arXiv:2506.13858.
%20
\bibitem{coldcollapse1999} B. Moore, T. Quinn, F. Governato, J. Stadel, G. Lake, Cold collapse and the core catastrophe, Mon. Not. R. Astron. Soc. \textbf{310}, 1147 (1999).
%21
\bibitem{McGetal24} S.S. McGaugh, J.M. Schombert, F. Lelli, and J. Franck, Accelerated structure formation: the early emergence of massive galaxies and clusters of galaxies, Astrophys. J. \textbf{976}, 13 (2024).
%22
\bibitem{Sil25} O.K. Sil'chenko, Galaxies in the first billion years of the Universe's expansion, Phys. Usp. \textbf{68}, 177 (2025).
%23
\bibitem{FerPalDay23} A. Ferrara, A. Pallottini, P. Dayal, On the stunning abundance of super-early, luminous galaxies revealed by JWST, Mon. Not. R. Astron. Soc. \textbf{522}, 3986 (2023).
%24
\bibitem{Blaetal25} M. Blamart, A. Liu, R. Brandenberger, J.B. Mu\~{n}oz, B. Cyr, UV luminosity functions from HST and JWST: a possible resolution to the high-redshift galaxy abundance puzzle and implications for cosmic strings, arXiv:2512.09980.
%25
\bibitem{Peretal25} P.G. Perez-Gonz\'alez, et al., The rise of the galactic empire: ultraviolet luminosity functions at $z\sim17$ and $z\sim25$ estimated with the MIDIS+NGDEEP ultra-deep JWST/NIRCam data set, Astrophys. J. \textbf{991}, 179 (2025).
%26
\bibitem{Aaretal25} L.Y.A. Yung, R.S. Somerville, K.G. Iyer, $\Lambda$CDM is still not broken: empirical constraints on the star formation efficiency at $z\sim12-30$, Mon. Not. R. Astron. Soc. \textbf{}, 3802 (2025).
%27
\bibitem{Bonetal91} J.R. Bond, S. Cole, G. Efstathiou, N. Kaiser, Excursion set mass functions for hierarchical Gaussian fluctuations, Astrophys. J. \textbf{379}, 440 (1991).
%28
\bibitem{LacCol93} C. Lacey and S. Cole, Merger rates in hierarchical models of galaxy formation, Mon. Not. R. Astron. Soc. \textbf{262}, 627 (1993).
%29
\bibitem{BerDokEro03} V. Berezinsky, V. Dokuchaev, Yu. Eroshenko, Small-scale clumps in the galactic halo and dark matter annihilation, Phys. Rev. D \textbf{68}, 103003 (2003).
%30
\bibitem{BodOstTur01} P. Bode, J. Ostriker, N. Turok, Halo formation in warm dark matter models, Astrophys. J. \textbf{556}, 93 (2001).
%31
\bibitem{ST} R.K. Sheth, G. Tormen, Large-scale bias and the peak background split, Mon. Not. R. Astron. Soc. \textbf{308}, 119 (1999).
%32
\bibitem{WhiRee78} S.D.M. White, M.J. Rees, Core condensation in heavy halos: a two-stage theory for galaxy formation and clustering, Mon. Not. R. Astron. Soc. \textbf{183}, 341 (1978).
%33
\bibitem{Fur06} S. Furlanetto, The global 21-centimeter background from high redshifts, Mon. Not. Roy. Astron. Soc. \textbf{371}, 867 (2006).
%34
\bibitem{Furlanetto2006}
S.R. Furlanetto, S.P. Oh, and F.H. Briggs, Cosmology at low frequencies: the 21 cm transition and the high-redshift Universe, Physics Reports \textbf{433}, 181 (2006).
%35
\bibitem{Meletal06} G. Mellema, I.T. Iliev, U.-L. Pen, P.R. Shapiro, Simulating cosmic reionization at large scales - II. The 21-cm emission features and statistical signals, Mon. Not. R. Astron. Soc. \textbf{372}, 679 (2006).
%36
\bibitem{Steetal08} K.R. Stewart, J.S. Bullock, R.H. Wechsler, A.H. Maller, A.R. Zentner, Merger histories of galaxy halos and implications for disk survival, Astrophys. J. \textbf{683}, 597 (2008).
%37
\bibitem{Genetal09} S. Genel, R. Genzel, N. Bouch\'e, T. Naab, A. Sternberg, The halo merger rate in the Millennium simulation and implications for observed galaxy merger fractions, Astrophys. J. \textbf{701}, 2002 (2009).
%38
\bibitem{TraCenMan15} H. Trac, R. Cen, P. Mansfield, SCORCH I: the galaxy-halo connection in the first billion years, Astrophys. J. \textbf{813}, 54 (2015).
%39
\bibitem{SomKol99} R.S. Somerville, T.S. Kolatt, How to plant a merger tree, Mon. Not. R. Astron. Soc. \textbf{305}, 1 (1999).
%40
\bibitem{NeiBosDek06} E. Neistein, F.C. van den Bosch, A. Dekel, Natural downsizing in hierarchical galaxy formation, Mon. Not. R. Astron. Soc. \textbf{372}, 933 (2006).
%41
\bibitem{NeiDek08} E. Neistein, A. Dekel, Merger rates of dark matter haloes, Mon. Not. R. Astron. Soc. \textbf{388}, 1792 (2008).
%42
\bibitem{MorGioShe08} J. Moreno, C. Giocoli, R.K. Sheth, Merger history trees of dark matter haloes in moving barrier models, Mon. Not. R. Astron. Soc. \textbf{391}, 1729 (2008).
%43
\bibitem{ParColHel08} H. Parkinson, S. Cole, J. Helly, Generating dark matter halo merger trees, Mon. Not. R. Astron. Soc. \textbf{383}, 557 (2008).
%44
\bibitem{Nadetal23} E.O. Nadler, A. Benson, T. Driskell, X. Du, V. Gluscevic, Growing the first galaxies' merger trees, Mon. Not. R. Astron. Soc. \textbf{521}, 3201 (2023).
%45
\bibitem{Deketal09} A. Dekel, Y. Birnboim, G. Engel, J. Freundlich, T. Goerdt, M. Mumcuoglu, E. Neistein, C. Pichon, R. Teyssier, E. Zinger, Cold streams in early massive hot haloes as the main mode of galaxy formation, Nature \textbf{457}, 451 (2009).
%46
\bibitem{StiRic26} M. Stiavelli, M. Ricotti, How bursty is star formation at $z>5$?, arXiv:2602.16706.
%47
\bibitem{Pilipenko26} S. Pilipenko, G. Yepes, S. Gottl{\"o}ber, S. Knollmann, Ginnungagap --- A massively parallel cosmological initial conditions generator, Astron. Comput. \textbf{55}, 101082 (2026).
%48
\bibitem{Springel05} V. Springel, The cosmological simulation code GADGET-2, Mon. Not. R. Astron. Soc. \textbf{364}, 1105 (2005).
%49
\bibitem{Behroozi13a} P.S. Behroozi, R.S. Wechsler, H.-Y. Wu, The ROCKSTAR phase-space temporal halo finder and the velocity offsets of cluster cores, Astrophys. J. \textbf{762}, 109 (2013).
%50
\bibitem{Behroozi13b} P.S. Behroozi, R.S. Wechsler, H.-Y. Wu et al., Gravitationally consistent halo catalogs and merger trees for precision cosmology, Astrophys. J. \textbf{763}, 18 (2013).
%51
\bibitem{Zhuetal26} H. Zhu, B. Yue, Y. Xu, X. Chen, and Z. Huang, Probing power spectrum enhancement at small scales with the SKA, Astrophys. J. \textbf{1001}, 25 (2026).
%52
\bibitem{MirFurSun17} J. Mirocha, S.R. Furlanetto, G. Sun, The global 21-cm signal in the context of the high-$z$ galaxy luminosity function, Mon. Not. R. Astron. Soc. \textbf{464}, 1365 (2017).
%53
\bibitem{Wecetal02} R.H. Wechsler, J.S. Bullock, J.R. Primack, A.V. Kravtsov, A. Dekel, Concentrations of dark halos from their assembly histories, Astrophys. J. \textbf{568}, 52 (2002).
%54
\bibitem{Deketal13} A. Dekel, A. Zolotov, D. Tweed, M. Cacciato, D. Ceverino, J.R. Primack, Toy models for galaxy formation versus simulations, Mon. Not. R. Astron. Soc. \textbf{435}, 999 (2013).
%55
%\bibitem{PriLoe12} J.R. Pritchard, A. Loeb, 21 cm cosmology in the 21st century, Reports on Progress in Physics \textbf{75}, 086901 (2012).
%56
%\bibitem{KraMan22} A. Kravtsov, V. Manwadkar, GRUMPY: a simple framework for realistic forward modelling of dwarf galaxies, Mon. Not. R. Astron. Soc. \textbf{514}, 2667 (2022).

\end{thebibliography}
\end{document}